\documentclass[%
reprint,
amsmath,amssymb,
prb,
]{revtex4-2}
\usepackage[english]{babel}
\usepackage{graphicx}% Include figure files
\usepackage{dcolumn}% Align table columns on decimal point
\usepackage{bm}% bold math
\usepackage{hyperref}% add hypertext capabilities
\usepackage{mathrsfs}
\usepackage{caption}
\usepackage{subcaption}
\usepackage{color}

\newcommand{\highlight}[1]{\textcolor{black}{#1}}

\begin{document}
	
	\preprint{APS/123-QED}
	
	\title{Realistic modeling of transport properties at finite temperature in magnetic materials by local quantization of a Heisenberg model}
	
	\author{Fabian Engelke}
	\author{Christian Heiliger}%
	\email{christian.heiliger@physik.uni-giessen.de}
	\affiliation{%
		Institute for theoretical Physics and Center for Materials Research Justus Liebig University Giessen \\
		Heinrich-Buff-Ring 16, 35392 Giessen, Germany
	}%
	
	\date{\today}
	
	\begin{abstract}
		The quantitative description of the electrical resistivity of a magnetic material remains challenging to this day. Qualitatively, it is well understood that the temperature-induced lattice and spin disorder determines the temperature dependence of the resistivity. While prior publications reached good agreement with experiment in the so-called supercell or direct approach for non-magnetic materials, where the spin-disorder contribution to the resistivity is negligible, an accurate, purely theoretical description of magnetic materials remains elusive. This shortcoming can be attributed to the missing accuracy in the description of the temperature-dependent spin-disorder itself. In this work, we employ a joint approach from \textit{ab-initio} transport calculations and atomistic modeling of the temperature-dependent spin-disorder. Using the example of $\alpha$-Fe, we demonstrate that including quantum-mechanical effects via a semiclassical local quantization of the Heisenberg model significantly improves the description of the spin-disorder component of the electrical resistivity. Compared to previous approaches, this model includes the description of magnetic short-range order effects, enabling us to study temperature effects around and above the Curie temperature, where prior mean-field theory-based approaches inevitably predicted a constant contribution.
	\end{abstract}

\maketitle

\section{Introduction}
Even though the origin of the temperature dependence of a material's electrical resistivity has long been understood qualitatively \cite{ziman_electrons_2007}, quantitative predictions without any input from experiment remain challenging. Qualitatively, it is well established that for most solids, temperature dependence is dominated by electron-phonon scattering. However, in magnetic systems, electron-magnon scattering, i.e., scattering of electrons at the disorder of the magnetic system, is important and may even become the dominant contribution \cite{weiss_spin-dependence_1959}. \par
Thus, an accurate description of the spin-disorder contribution, i.e., the spin-disorder resistivity (SDR), is essential to research in spintronics. Further, on a more fundamental level, the temperature dependence of electrical resistivity may be considered a rare probe to study the nature of spin fluctuations in solids. \par
Classically, the experimentally observed electrical resistivity has been decomposed into an electron-phonon and electron-magnon contribution under the assumption of independent contributions, i.e., the validity of Matthiessen's rule, and a saturating spin-disorder contribution way above the Curie temperature \cite{weiss_spin-dependence_1959}. However, more recent investigations could demonstrate deviations from Matthiessen's rule, especially for rare-earth elements \cite{glasbrenner_deviations_2014}. For $\alpha$-iron, deviations from Matthiessen's rule are much smaller and remain small compared to the total electrical resistance \cite{liu_direct_2015,glasbrenner_deviations_2014}. Nonetheless, the isolated description of spin-disorder resistivity, as presented in this work, should be understood as a first step toward a holistic description of electrical resistivity, which would combine lattice and spin-disorder effects. \par
In the past, spin-disorder resistivity has been extensively studied using an \textit{sd}-Hamiltonian. The solid is modeled by a set of $s$-like electrons, which carry the majority of an electric current, and localized \textit{d}-orbitals carrying the magnetic moment. The spin-direction-dependent exchange interaction between \textit{s} and \textit{d} electrons then leads to a temperature dependence of the electrical resistivity \cite{kasuya_electrical_1956}. Note that magnetic short-range order effects have also been investigated in the framework of an \textit{sd}-Hamiltonian \cite{kataoka_resistivity_2001,akabli_effects_2008}. Even though the model provides a physical explanation for the emergent spin-disorder resistivity, many of the model's core assumptions are questionable. For instance, Goodings showed that scattering is dominated by \textit{s-d} transitions \cite{Goodings1963-nw}, questioning the clear separation into current-carrying s-electrons and localized \textit{d}-electrons. \par
More recently, spin-disorder resistivity in the paramagnetic state has been modeled utilizing a so-called disordered local moments (DLM) state. Spin-disorder in that approach is modeled by creating an effective medium equivalent to a state with uncorrelated randomly oriented spins by means of the coherent-potential approximation \cite{kudrnovsky_spin-disorder_2012,glasbrenner_deviations_2014,drchal_spin-disorder_2018}. This alloy analogy can be extended to represent states of arbitrary magnetization, establishing a temperature dependence by mapping over the (experimental) magnetization curve \cite{Ebert2015-ay}. The electrical resistivity corresponding to such a DLM state is then typically determined in the Kubo-Greenwood formalism \cite{kubo_statistical-mechanical_1957,greenwood_boltzmann_1958,Ebert2015-ay}. \par
An alternative approach involves the simulation of thermal-induced disorder in large supercells and the determination of electric properties in the framework of the Landau-Büttiker formalism \cite{datta_electronic_2009}. Note that, as opposed to the DLM approach, this description allows the consideration of the specific magnetic short-range order in the system. Wysocki \textit{et al.} \@\cite{wysocki_first-principles_2009} used the mean-field approximation to create magnetization-dependent configurations of spin disorder, once again only allowing for the introduction of temperature dependence by mapping over the Curie-curve. Liu \textit{et al.} \@\cite{liu_direct_2015} establish a direct temperature dependence by creating the spin disorder by superposition of magnon modes. However, in this procedure, magnons are modeled as massless, leading to a significant overestimation of the spontaneous magnetization and, consequently, the general magnetic order. This results in an underestimation of electrical resistivity at higher temperatures ($T > 100 K$). Once again, the remedy is to rescale the temperature using the experimental magnetization curve. \par
Besides the dependence on experimental data, the temperature rescaling based on magnetization has one further shortcoming: the spin-disorder resistivity is inevitably constant above the Curie temperature, as vanishing magnetization forbids any magnetization-based rescaling. Nonetheless, as we will demonstrate below, the magnetic short-range order will continue to decrease above the Curie point, which in turn will lead to an increasing spin-disorder component in the electrical resistivity. \par
A suitable approach to modeling temperature-induced spin-disorder is the classical Heisenberg model \cite{halilov_magnon_1997,turek_exchange_2006}, which offers a reasonable description of the phase transition itself and properties in the paramagnetic state. Yet, due to its purely classical nature, it fails to describe disorder in the ferromagnetic state accurately \cite{evans_quantitative_2015,kormann_rescaled_2010, kormann_role_2011}. Thus, the inclusion of quantum-mechanical effects is inevitable. However, a full quantum mechanical treatment by means of quantum Monte Carlo (QMC) techniques \cite{sandvik_computational_2010} is computationally infeasible for the large supercells required in transport calculations and encounters the sign-problem, which describes instabilities arising in systems with competing interactions \cite{henelius_sign_2000}. More recently, Walsh \textit{et al.}\@ \cite{walsh_realistic_2022} showed that one can effectively describe quantization effects by a local spin-quantization technique, allowing for a numerically efficient simulation of spin-disorder including quantization effects. \par
In the following section, we briefly review the main ideas of this approach before outlining the details of our supercell-based methodology employed for determining electrical resistivity. In Section \ref{sec:results}, we apply this approach to $\alpha$-iron and present our findings for the effect of quantization on the magnetic order in the system. Based on the simulation of the temperature-induced magnetic disorder, we continue to obtain the spin-disorder component of the electrical resistivity, demonstrating that the Heisenberg model, in general, is a suitable platform for the description of spin-disorder effects on electronic transport properties. Additionally, we show that the magnetic short-range order influences the resistivity in the paramagnetic state, and that including quantum mechanical effects significantly improves the description in the ferromagnetic state.
\section{Methods}
To determine the spin-disorder component of the electrical resistivity, we employ the direct or supercell approach, leveraging the computationally efficient description of temperature-induced spin disorder by atomistic simulations and the high accuracy of non-collinear density functional theory (DFT) to extract the system's temperature-dependent electron transport properties. Note that this approach allows us to study the influence of magnetic short-range order on the resistivity. \par
In the following, we first describe the methodology of the atomistic simulations and then our DFT calculations.
\subsection{Magnetic Disorder}
The first and arguably key step in the determination of the temperature-dependent transport properties is an accurate description of the temperature-induced disorder of the magnetic system. It is well established that $\alpha$-Fe is well-described by localized atomistic spins, governed by a Heisenberg Hamiltonian of the form:
\begin{equation}
	\mathscr{H} = -\sum\limits_{\langle ij \rangle} J_{ij}\vec{S}_i\vec{S}_j
	\label{eq:heis_hamilton}
\end{equation}
\highlight{where $\vec{S}_i$ describe a spin on site $i$ of a \textit{bcc}-lattice in the case of $\alpha$-iron.} Note that other interactions complementing the isotropic exchange interaction, such as anisotropy energy, Dzyaloshinskii Moriya interactions, or dipole interactions, could be included in the model but are negligible in iron at the length scales and temperatures considered in this work \cite{evans_atomistic_2014}. Further, the local magnetic moment in iron is found to be fairly stable with respect to temperature, which is why longitudinal spin fluctuations can safely be neglected for this system \cite{ruban_temperature-induced_2007}.  \par
\begin{figure}
	\begin{subfigure}{0.21\textwidth}
		\includegraphics[width=\textwidth]{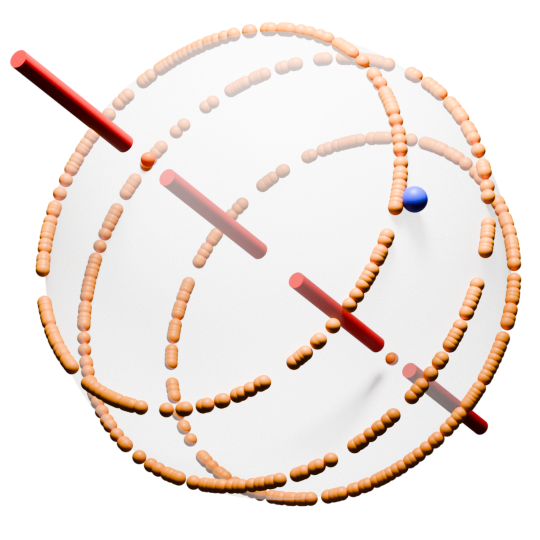}
		\caption{SMC}
	\end{subfigure}
	\begin{subfigure}{0.21\textwidth}
		\includegraphics[width=\textwidth]{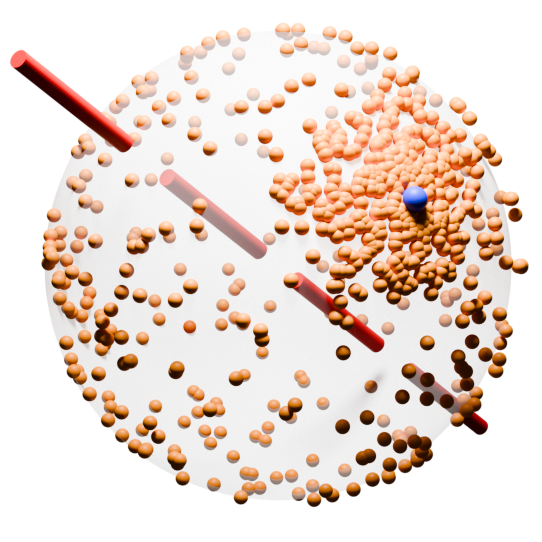}
		\caption{CMC}
	\end{subfigure}
	\caption{Visualization of trial-moves in Monte Carlo simulations in the classical case (CMC) and semiclassically quantized case corresponding to a spin quantum number of $s=2$ (SMC) for a given local magnetic field (red dashed axis) and initial spin (blue dot). Classical simulations are performed employing the Hinzke-Nowak strategy for optimized sampling. Visualized trial moves correspond to a temperature of 100~K.}
	\label{fig:trial_spin_visualization}
\end{figure}

To obtain a temperature-dependent spin configuration corresponding to this Hamiltonian, one can either use Langevin dynamics in the time integration of the Landau-Lifshitz Gilbert equation \cite{garcia-palacios_langevin-dynamics_1998} or, equivalently, perform Monte Carlo simulations in the standard Metropolis algorithm \cite{metropolis_equation_1953,landauGuideMonteCarlo2015}. Optimized sampling methods may be employed to ensure efficient phase space coverage. That is, we do not use uniformly distributed trial spins in the Monte Carlo simulation but instead use the Hinzke-Nowak algorithm, which increases the acceptance rate while ensuring ergodicity \cite{hinzke_monte_1999} (cf. \@ figure \ref{fig:trial_spin_visualization}). \par
The classical model, i.e., treating the spins continuously, allows a reasonable description of the phase transition. However, the lack of quantum mechanical effects leads to a poor description of the system in the ferromagnetic phase. Thus, to obtain an accurate description for all temperatures, the classical spins $\vec{S}_i$ have to be promoted to quantum operators. \par
Avoiding the high computational cost and instabilities in QMC, we employ the semiclassical local quantization scheme proposed by Walsh \textit{et al.}\@ \cite{walsh_realistic_2022} in this work. The simplifications made are based on the following observations: At first, in most magnets, entanglement is only important at temperatures up to a few degrees Kelvin \cite{amico_entanglement_2008}, which is why the wave function of the whole system might be approximated as a product state. The second crucial step is to use a local quantization scheme. In a classical Monte-Carlo (CMC) simulation, a spin in the system is chosen at random, and a new direction of the chosen spin is proposed. This new spin can, in principle, take every direction on the unit sphere. The idea of semiclassical local quantization (SMC) is to treat the proposition, and just the proposition of trial spins quantum mechanically, i.e., only allow spins quantized in the direction of the local magnetic field given by:
\begin{equation}
	\vec{B}_i = -\dfrac{\partial\mathscr{H}}{\partial \vec{S}_i}
\end{equation}  \par 
Figure \ref{fig:trial_spin_visualization} illustrates trial spins for a classical description using the Hinzke-Nowak algorithm for optimized sampling \cite{hinzke_monte_1999} employed in this work and the semiclassical quantized trial moves. \par
Obviously, the choice of the spin quantum number $s$, characterizing the local magnetic moments, is crucial for the model. Note that the classical case is recovered for $s\rightarrow\infty$, but $s=1/2$ does not correspond to the Ising model, as the spin direction is chosen uniformly in the case of a vanishing local magnetic field. However, despite its influence, how $s$ should be determined for a practical calculation is non-trivial. \highlight{This is, in particular, as the Heisenberg model is only an effective description of the magnetic ordering, and there is no direct link to the experimental electronic structure. Thus, the spin quantum number is a bear model parameter at first, determined to describe the quantization properties experienced by the magnetic density in the solid.} Here, we adhere to the established methodology of directly correlating $s$ to the magnitude of the local magnetic moment. \par
As a further consequence of the transition from classical to quantized spins, one must rescale the exchange parameter $J_{ij}$ in Eq.\@  (\ref{eq:heis_hamilton}). Taking the limit of infinite quantum spin number $s$, one finds the following scaling relationship:
\begin{equation}
	s(s+1)J_{ij}^\text{q} = s^2J_{ij}^\text{c}
\end{equation}
Note that the introduction of quantum mechanical effects might also be realized without the specific quantization of spins. Woo \textit{et al.}\@ \cite{woo_quantum_2015} showed that such effects might be introduced by imposing a Bose-Einstein distribution for occupied magnon-modes by connecting the spin system to a quantum heat bath, leading to a rescaled temperature. However, this approach (in the quasi-harmonic approximation \cite{woo_quantum_2015}) relies on prior knowledge of the system's Curie temperature and reverting to the classical description in the paramagnetic state. \par
The model employed here is free of such an artificial transition and naturally coincides with the classical description for magnetization as well as other thermodynamic properties, especially the electric resistivity, as we will demonstrate below. \par
\subsection{Transport}
\begin{figure}
	\includegraphics[width=0.45\textwidth]{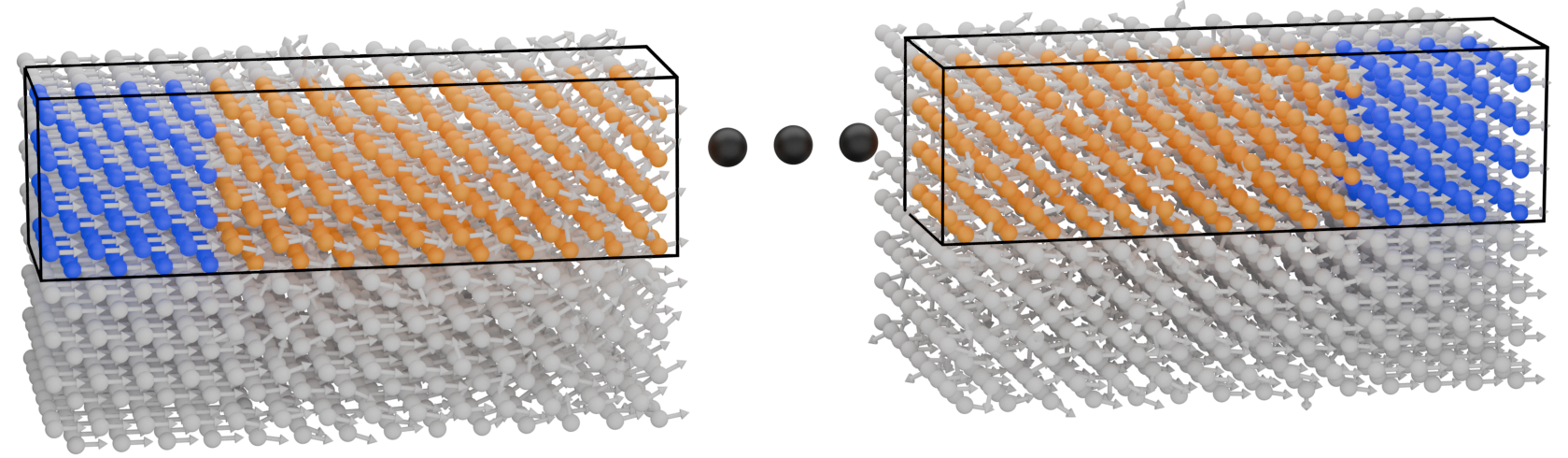}
	\caption{Schematic representation of the transport geometry employed in the calculations. Lead atoms (blue) are infinitely repeated in positive and negative transport directions. The length of the scattering region, i.e., the thermally disordered material (orange), is modulated to extract the specific resistivity without any effects from the leads. The used $4\times 4$ in-plane supercells are periodically continued (gray).}
	\label{fig:transport_geometry}
\end{figure}
\highlight{In order to determine the specific resistivity, we employ the so-called supercell or direct approach. The basis of such a calculation is formed by the calculation of the electrical conductance, or equivalently resistivity, in an ensemble of magnetically disordered supercells of varying length. This requires the simulation of electronic transport in a realistic transport geometry. In our case, we connect the disordered supercell with leads formed by collinear iron. These leads are perfect conductors and are formally extended to positive(negative) infinity in transport direction (cf.\@ figure 2). We subsequently employ the Non-equilibrium Green's function formalism in a Korringa-Kohn-Rostoker \cite{korringa_calculation_1947,kohn_solution_1954} representation (NEGF-KKR), which, as implemented in the in-house Giessen KKR code \cite{heiliger_implementation_2008}, allows for the consideration of atomically resolved non-collinear spin structures \cite{czerner_fully_2008}, to obtain the transmission function $T(E)$. The famous Landauer formula then gives the connection to the electrical current originating from the left lead and entering the right lead \cite{landauerConductanceTransmissionCommon1992,datta_electronic_2009}:
\begin{equation}
	j_\text{LR} =  \frac{e}{h}\int \mathrm{d}E \; T(E)\left( f_\text{L}(E)-f_\text{R}(E)\right)
\end{equation}
where $f_\mathrm{L/R}(E)$ are the distribution functions in the left (right) lead. Considering the limit of a vanishing bias voltage, the electrical resistivity may be obtained by:
\begin{equation}
    (RA)^{-1} = \frac{G_0}{2}\int \mathrm{d}E\; T(E)\partial_Ef(E,T)
\end{equation}
where $A$ is the cross section of the supercell, $f(E,T)$ is the Fermi-Dirac distribution function, and $G_0$ is the conductance quantum. Note that this formulation includes elastic scattering only. However, as we consider a vanishing bias voltage, i.e., we consider the linear response, the elastic contribution dominates the electronic transport properties.}
Further, we restrict the evaluation of this integral to a single energy point at the Fermi level of the collinear system. As we observe a nearly antisymmetric behavior of the transmission function around the collinear Fermi energy, we expect good results even at higher temperatures due to error cancellation.
\highlight{In the Ohmic transport regime, i.e., sufficient scattering occurs over the length of the supercell, the overall resistance of the material will scale linearly with the length of the device $z$, i.e., $RA\propto \rho z$. Thus, the specific resistivity $\rho$ characteristic of the scattering region may be extracted from the length dependence of the resistance area product, by means of a linear fit.} \par
%\textcolor{red}{For the determination of the specific electrical resistivity, we employ the so-called supercell approach. At its core}, we use the Non-equilibrium Green's function technique in a Korringa–Kohn–Rostoker representation \cite{korringa_calculation_1947,kohn_solution_1954,heiliger_implementation_2008}, allowing the consideration of non-collinear spin-structures \cite{czerner_fully_2008} implemented in the Giessen in-house KKR-code, \textcolor{red}{to obtain the transmission function, then related to the conductance of the supercell by the Landauer formula}. This formalism allows the calculation of the conductance in a realistic transport geometry, e.g., a disordered supercell sandwiched by perfectly collinear leads. This geometry is schematically depicted in figure \ref{fig:transport_geometry}. \par
%In the Ohmic transport regime, the overall resistance of the material will scale linearly with the length of the device. Thus, the specific resistivity characteristic of the scattering region may be extracted from the length dependence of the resistance area product, \textcolor{red}{i.e., the reciprocal conductance given by the Landauer Formula}. Thus, we perform conductance calculations on an ensemble of supercells of varying lengths to extract the material's specific resistivity. \par
To avoid self-interaction effects, one has to perform atomistic simulations of the thermal spin disorder in large unit cells. We find that cells spanning at least $20\times 20 \times 20$ \textit{bcc} cells show well-converged results. As these cells are far too large for treatment in a DFT-based transport calculation, we must cut smaller supercells from those used for the atomistic simulation. Note that the spin disorder in the transport supercell will not fulfill the in-plane periodic boundary conditions imposed by our transport calculations. Thus, the periodic continuation of the cell in the DFT calculation will introduce planes of artificially low magnetic short-range order. However, we find well-converged results for in-plane supercells of $4\times 4$ \textit{bcc} unit cells. \par
In the setup of a single NEGF-KKR calculation, we use the atomic-sphere approximation. All calculations are based on a self-consistent potential, which includes relativistic effects. A \textit{spdfgh} basis set in the angular momentum expansion is employed for the self-consistent calculation. The transport calculations are one-shot using the potential of the collinear phase. In order to reduce computational cost of the transport calculations, the basis set for Green's function expansion is truncated to \textit{spd}, and we neglect relativistic effects. Note that the transport properties are solely determined by scattering at the potentials and not the electron density. Thus, the improved accuracy of the self-consistent potential calculation, through relativistic effects and the larger basis set, improves the accuracy of the scheme compared to consistent but less accurate numerical parameters in both calculation steps. \par
\section{Results}
\label{sec:results}
\subsection{Thermal Magnetic Disorder}
The exchange parameters $J_{ij}$ used in this work are taken from Pajda \textit{et al.}\@ \cite{pajda_ab_2001}, who obtained the parameters from \textit{ab-initio} calculations. We consider exchange interactions between the first 10 shells, omitting the $(1/2,1/2,5/2)$ position in the 10th shell, which shows much weaker exchange interactions than the $(3/2,3/2,3/2)$ position \cite{antropov_aspects_1999}. \par
The choice of the spin quantum number $s$ is pivotal for the semiclassical quantization scheme. Focusing on the effective character of the localized atomistic spin model, one might neglect the complex origin of the local magnetic moments and focus solely on the magnitude. For iron, our DFT calculations predict a local magnetic moment of $2.26\mu_\text{B}$, which stands in good agreement with the experimental value of $2.22 \mu_\text{B}$ \cite{Kittel2004-lf}. \highlight{Under the assumption, that the Landé factor $g=2$ for $\alpha$-iron, which is justified given the weak spin orbit coupling, and also supported by experimental results \cite{jonesAccurateDeterminationElectron2022}}, this implies a spin quantum number of $s=1.1$, suggesting running simulations with $s=1$ or $s=3/2$, or even interpolating between both as has been done in previous studies \cite{kormann_rescaled_2010}.  \par
\begin{table}
	\caption{\label{tab:curie_temp}%
		Comparison of Curie temperatures obtained from Binder cumulant crossing analysis in the proposed MC schemes and the experimental value.
	}
	\begin{ruledtabular}
		\begin{tabular}{lcccr}
			&
			\textrm{CMC}&
			\textrm{SMC(s=1)}&
			\textrm{SMC(s=3/2)} &
			Exp. \cite{crangle_magnetization_1971} \\
			\colrule
			$T_\text{C}$ [$K$] &  1050 &  1097 &  1086 & 1044 \\
		\end{tabular}
	\end{ruledtabular}
\end{table}
A key quantity for characterizing any ferromagnetic material is the Curie temperature. \highlight{Here, we determine the critical temperature of the phase transition from a Binder-cumulant crossing analysis, in order to avoid finite size effects. That is, we calculate the Binder cumulant
\begin{equation}
    U_4 = \frac{3}{2}\left( 1-\frac{\langle m^4 \rangle}{3\langle m^2\rangle^2}\right)
\end{equation}
for cubic spin cells with an edge length of $L=4,8,12,16,20,24$ \textit{bcc} unit cells, in 1~K intervals around the Curie temperature over 300,000 Monte Carlo passes, following 30,000 equilibration steps. The value of $U_4$ is independent of the system size at the Curie point alone -- given a large enough system -- allowing one to obtain the critical temperature from the crossing point of Binder cumulants $U_4(L)$ corresponding to different system sizes \cite{binderFiniteSizeScaling1981,binderMonteCarloSimulation2010}. We employ a smoothing cubic spline for the determination of the crossing point of two cumulants.} \par
\highlight{Further, the finite-size error $\Delta T_\text{C}$ of the Curie temperature, obtained from the crossing point of $U_4(L), U_4(L^\prime)$ is expected to scale as \cite{binderFiniteSizeScaling1981}:
\begin{equation}
    \Delta T_\text{C}\propto \ln \left( \frac{L^\prime}{L} \right)^{-1}
\end{equation}
Thus, we calculate the crossing points of $L$ with all available $L^\prime>L$ and estimate the Curie temperature for each $L$ by linearly extrapolating the crossing points $(L,L^\prime)$ to $\ln \left( {L^\prime}/{L} \right)^{-1}\rightarrow 0$. We find good agreement between the transition temperatures obtained by extrapolating the data for $L=8,12,16$, with slight deviations for $L=4$ across all considered Monte Carlo schemes.} \par
Table \ref{tab:curie_temp} lists the results based on the averaged estimates from $L=8,12,16$. The CMC result stands in excellent agreement with the experimental value, while the SMC results show a slight overestimation. \par
%A key quantity for characterizing any ferromagnetic material is the Curie temperature. Here, we employ the fluctuation dissipation theorem to obtain the temperature-dependent specific heat capacity of the system at hand and determine the Curie temperature by fitting the peak in the spectrum. Table \ref{tab:curie_temp} depicts the results for CMC and SMC simulations. The CMC result stands in excellent agreement with the experimental value, while the SMC results show a slight overestimation. \par
\begin{figure}
	\includegraphics[width=0.5\textwidth]{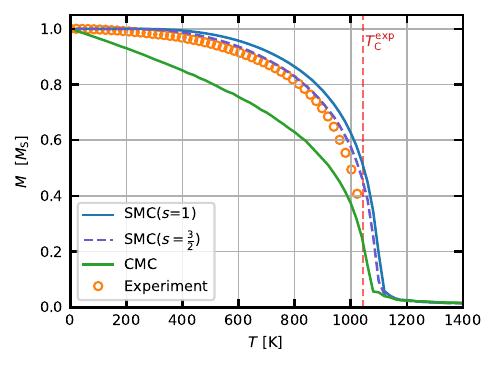}
	\caption{Temperature-dependent reduced magnetization for SMC and CMC simulations compared to experimental values taken from ref.\@ \cite{crangle_magnetization_1971}. MC simulations were performed for $22\times 22\times 22$ \textit{bcc} unit cells. 50,000 equilibration passes are performed before averaging the magnetization over 1,000 spin-cell passes.}
	\label{fig:curie_curve}
\end{figure}
Figure \ref{fig:curie_curve} depicts the obtained temperature-dependent magnetization for CMC and SMC simulations. Reproducing the results obtained by Walsh \textit{et al.}\@ \cite{walsh_realistic_2022}, agreement with the experimental values is significantly improved for SMC compared to CMC. We find better agreement for higher s values at low temperatures, even though magnetization is generally overestimated in SMC and greatly underestimated for CMC simulations. At temperatures closer to the Curie point, we find better agreement for $s=1$ than $s=3/2$. CMC continues to underestimate spontaneous magnetization. \par
\begin{figure}
	\includegraphics[width=0.5\textwidth]{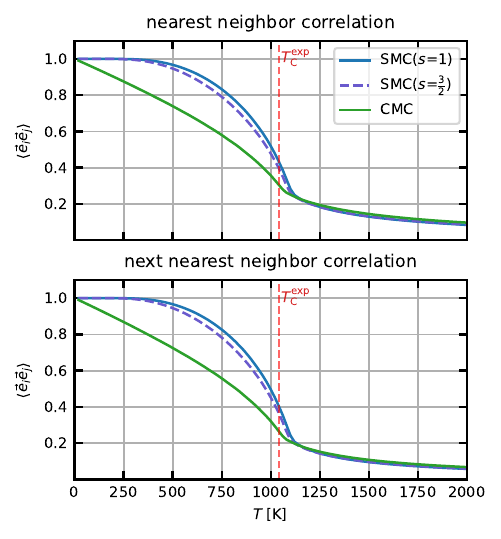}
	\caption{Temperature dependence of the normalized spin correlation functions for nearest and next nearest neighbors in CMC and SMC simulations. Correlators result from averaging scalar products over all spins in a $22\times 22\times 22$ \textit{bcc} cells supercell, equilibrated over 50,000 passes.}
	\label{fig:correlator}
\end{figure}
The apparent higher degree of order at low temperatures in SMC simulations compared to CMC is also reflected in the averaged pair-correlation functions (cf. \@figure \ref{fig:correlator}). Notably, all simulations converge to one another above the Curie temperature. Further, the spin-correlation function continuously decreases above the Curie temperature, indicating decreasing magnetic short-range order, even after the long-range order in the system has already vanished. \par
\begin{figure}
	\includegraphics[width=0.5\textwidth]{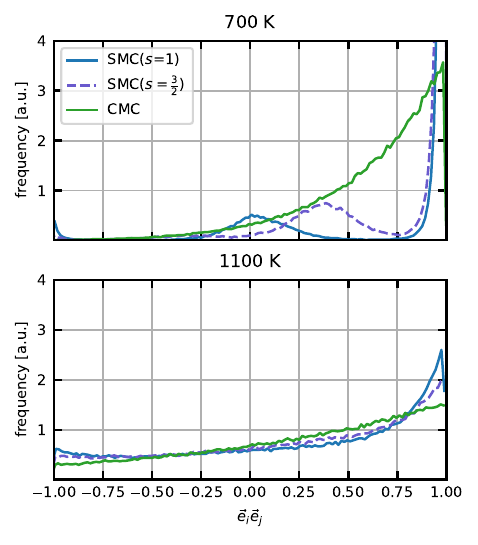}
	\caption{Distribution of nearest neighbor correlators below ($700$~K) and above ($1100$~K) the Curie temperature, for CMC and SMC simulations.}
	\label{fig:cor_dist}
\end{figure}
The convergence of SMC to CMC above the Curie temperature can also be observed in the nature of the disorder itself. Analyzing the distribution of correlators in SMC and CMC simulations (cf. \@figure \ref{fig:cor_dist}), we see that for temperatures below the Curie temperature, the probability for two neighboring spins being nearly parallel is far higher in SMC than in CMC calculations. The distribution of correlators experiences clear local maxima at values corresponding to the secondary angular momentum quantum numbers. Opposed to this, the frequency of correlators behaves monotonously for CMC calculations. From this, we conclude that disorder in SMC simulated systems is driven by point-like defects embedded in a nearly collinear spin system, whereas CMC systems exhibit more homogeneous disorder. However, as for the averaged correlators, this difference deteriorates for higher temperatures, especially for temperatures past the Curie point, and the system homogenizes in both simulations. \par

The point-like disorder behavior might also explain the overestimation of magnetization at low temperatures. This behavior, i.e., an underestimation of excitations at low temperature, was also observed based on other thermodynamic properties \cite{walsh_realistic_2022}, suggesting a systematic under-representation of \highlight{long-wavelength} low-temperature excitations. \highlight{Reminiscent of this shortcoming is the behavior of the spontaneous magnetization in the low temperature limit. Instead of the expected $T^{3/2}$ (Bloch's Law \cite{blochZurTheorieFerromagnetismus1930}) or general power-law \cite{koblerEffectiveSpinQuantum2003} scaling relationship, one finds a rather asymptotic behavior of the magnetization.} \par
\highlight{Classically, low-temperature excitations are understood as spin-waves, described by an infinitesimal deviation of the spin's direction from the collinear arrangement. In SMC simulations, such small deviations only appear as a consequence of the changing magnetic field in the surroundings of a spin changing the quantization level. However, as the quantization level change is associated with a large energy penalty, such events are rare at low temperatures. Consequently, disorder only appears in the form of local cluster nucleating around such a defect, which negates the character of pure long-wavelength excitations.} \par
Further, considering that the surrounding spins can gain energy by tilting back into the direction of the local magnetic field, the net energy difference of an entire point defect will potentially be smaller than the activation energy of the seed alone. \highlight{Thus, the strictly local treatment of quantization may further suppress the formation of disorder, especially over longer length-scales. This may be remedied by the consideration of multiple spins at the same time. However, such analysis is subject to future work.}

\subsection{Resistivity of Iron}
We calculate the conductance in 10 different supercells with lengths between 25 and 55 \textit{bcc} unit cells to extract the specific resistivity. Further, we use the experimental lattice constant at $0$ K of $2.866$ $\text{\normalfont\AA}$ \cite{wilburnHydrostaticCompressionIron1978} for all calculations. \par
To ensure statistical independence of the used configurations, these supercells are taken from random non-overlapping positions in 24 independent MC simulations with a cell size of $22\times 22\times 55$ \textit{bcc} unit cells. In all cases, we find linear dependence of the resistivity area product on the scattering region length, indicating that transport is, in fact, ohmic and allowing us to determine the specific resistivity from the slope of the curve. \par
As the spin-disorder contribution to the electrical resistivity is not an observable on its own, a meaningful comparison to the experiment is not possible in a direct manner. Thus, we turn to results from Liu \textit{et al.} \@ \cite{liu_direct_2015} for the phonon contribution to the resistivity, allowing us to compare to the experiment under the assumption that Matthiessen's rule is valid and the resistivities are additive. For comparison in the entire temperature range, we extrapolate the results of Liu \textit{et al.} \@ \cite{liu_direct_2015} linearly as this is the theoretically expected, and computationally supported \cite{Verstraete2013-nh}, temperature dependence in the high-temperature limit for the electron-phonon contribution. \par
\begin{figure}
	\includegraphics[width=0.5\textwidth]{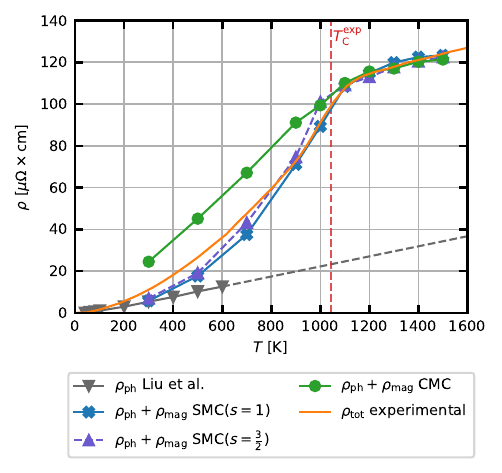}
	\caption{Comparison of total specific electrical resistivity based on CMC and SMC simulations with experimental data, under the assumption of Matthiessen's rule. Experimental data is taken from ref. \@ \cite{white_electrical_1959}.}
	\label{fig:rho_tot}
\end{figure}
\begin{table}
	\caption{\label{tab:sdr_comp}%
		Results for the specific resistivity in $\alpha$-iron based on spin-disorder alone.
	}
	\begin{ruledtabular}
		\begin{tabular}{lccr}
			\textrm{$T$ [K]}& & \textrm{$\rho_{\text{mag}}$ [$\mu\Omega\times \text{cm}$]} & \\
			&
			\textrm{CMC}&
			\textrm{SMC($s$=1)}&
			\textrm{SMC($s$=3/2)}\\
			\colrule
			300  &  19.15 &  0.43 &  1.55  \\
			500  &  34.95 &  7.49 &  8.91  \\
			700  &  52.11 & 22.52 & 28.17  \\
			900  &  71.30 & 54.75 & 51.26  \\
			1000 &  77.16 & 67.06 & 77.16  \\
			1100 &  85.43 & 83.77 & 84.36  \\
			1200 &  88.51 & 87.77 & 85.95  \\
			1300 &  86.16 &	90.40 & 88.20  \\
			1400 &  88.39 & 90.62 & 88.58  \\
			1500 &  87.05 & 89.02 & 88.22  \\
		\end{tabular}
	\end{ruledtabular}
\end{table}
Figure \ref{fig:rho_tot} depicts the results. We find excellent quantitative agreement in the paramagnetic phase for both the SMC and CMC simulations. As temperatures decrease, SMC shows the correct qualitative behavior, but the quantitative agreement worsens continuously. CMC fails to reproduce the experimentally observed qualitative trend in electrical resistivity. While CMC overestimates the electrical resistivity for all temperatures in the ferromagnetic phase, SMC underestimates the resistivity for low temperatures. Both may be traced back to the under (CMC) or overestimation (SMC) of magnetic order at low temperatures, which also surfaces in the under and overestimation of the magnetization. \par
\begin{figure}
	\includegraphics[width=0.5\textwidth]{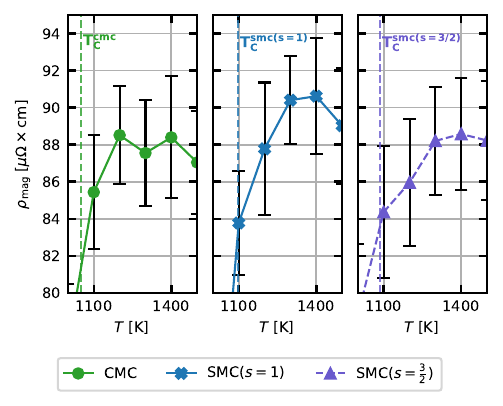}
	\caption{Comparison of isolated spin-disorder resistivity obtained from CMC and SMC-based calculations above the Curie temperature. Error bars indicate statistical uncertainty based on the fitting procedure alone.}
	\label{fig:sdr_comp}
\end{figure}
Looking at the isolated SDR contribution in the paramagnetic phase (cf. \@table \ref{tab:sdr_comp}, figure \ref{fig:sdr_comp}), an increase of resistivity after the Curie temperature up to roughly 1200~K in the CMC and 1300~K in the SMC simulation can be resolved, before the resistivity saturates. This increase in resistivity, which is more pronounced for SMC simulations, may be attributed to the decreasing magnetic short-range order and indicates a weak but non-negligible dependence of the electrical resistivity on the magnetic short-range order in the spin system. \par
\begin{figure}
	\includegraphics[width=0.5\textwidth]{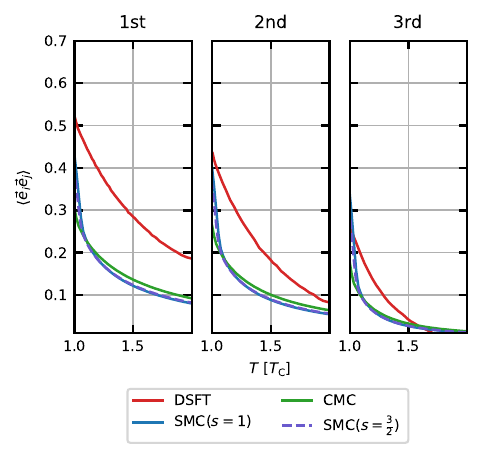}
	\caption{Comparison of normalized spin-correlation function of CMC, SMC, and results obtained from dynamical spin fluctuation theory \cite{melnikov_magnetic_2019} for the first three shells at temperatures above the Curie temperature. Temperatures are scaled to the Curie temperature because DSFT results significantly overestimate the experimental value.}
	\label{fig:dsft_comp}
\end{figure}

Previous studies assumed a significant underestimation of magnetic short-range order in the Heisenberg model. Most notably, results from dynamical spin fluctuations theory (DSFT) suggested higher values for nearest-neighbor correlation functions and thus a higher degree of magnetic short-range order \cite{melnikov_magnetic_2019}. Figure \ref{fig:dsft_comp} shows correlation functions for the first three shells in the Heisenberg model and the results obtained by Melnikov \textit{et al.} \@\cite{melnikov_magnetic_2019} using DSFT. \par
Considering the excellent agreement of the total resistivity for our calculation in the high-temperature limit ($T\gtrapprox T_\mathrm{C}$), one might argue that the description in the Heisenberg model is, in fact, correct, and DSFT overestimates magnetic short-range order. However, as the comparison with the experiment assumes the validity of Matthiessen's rule, which is an approximation \cite{glasbrenner_deviations_2014,liu_direct_2015}, the excellent agreement might be misleading. Previous studies have shown that the simultaneous consideration of lattice and spin disorder results in higher resistivities than the sum of their individual contributions. Thus, \highlight{even though we do not expect large deviations for $\alpha$-iron \cite{liu_direct_2015,glasbrenner_deviations_2014}}, the excellent agreement with the experiment might be due to error cancellation. \par
Further, Wysocki \textit{et al.}\@ demonstrated a clear dependence of the electrical resistivity on the local magnetic moment \cite{wysocki_first-principles_2009}. Our DFT calculations yield a local magnetic moment of 2.26$\mu_\text{B}$, slightly overestimating the experimentally observed magnetic moment of 2.22$\mu_\text{B}$. If anything, this leads to a slight overestimation of the resistivity in our calculations. The same can be said about the periodic continuation of supercells without periodicity, which leads to a decreasing magnetic short-range order in the DFT calculation. One further source of error in our approach is the missing self-consistency of transport calculations. The influence of this shortcoming on the final result remains unclear.
\section{Summary \& Outlook}
We propose using a semiclassical locally quantized Heisenberg model as the basis for a computationally feasible determination of temperature-dependent electron transport properties in magnetic materials. Including quantum mechanical effects in the model substantially improves the qualitative behavior in the ferromagnetic phase. Further, the temperature-based description of the disorder in the solid offers the opportunity to study magnetic short-range order effects, resolving the increasing spin-disorder contribution to the electrical resistivity above the Curie temperature in agreement with the experiment.  \par
Assuming the validity of Matthiessen's rule, we find excellent agreement with experimental values for the total electric resistivity in $\alpha$-iron. However, as past studies found deviations from Matthiessen's rule, further investigations, including lattice and spin disorder simultaneously, are necessary to conclude the accuracy of the magnetic short-range order description in the presented approach. Further, it might be possible to create spin cells of disorder corresponding to the correlation functions determined by DSFT via reverse Monte Carlo simulations. \par
In agreement with Walsh \textit{et al.}\@ \cite{walsh_realistic_2022}, the results for the electrical resistivity suggest a systematic under-representation of low-temperature excitations in the local quantization scheme employed. The effect of this shortcoming is amplified in the calculations for the electrical resistivity compared to the magnetization, where deviations from the experiment are smaller. The point-like character of disorder in the SMC simulations further supports the claim that the simultaneous quantization of multiple neighboring spins, as already suggested by Walsh \textit{et al.}\@\cite{walsh_realistic_2022}, improves the systematic under-representation of excitations at low temperatures, which would directly translate to an improved description of transport properties at low temperatures. \par
The determination of the used spin-quantum number remains unclear. The standard procedure, e.g., $s = M / (2\mu_\text{B})$, yields good results for $\alpha$-iron. However, it is unclear how this should be extended to materials with a smaller local magnetic moment. To avoid making $s$ to an undetermined fit-parameter in the theory, a framework that allows the determination of $s$ from first principles needs to be established. \par
Clearly, the problem of accurately predicting the spin-disorder contribution to electric resistivity is not solved. Nonetheless, the proposed scheme shows promising results in the high-temperature limit, where quantum effects are typically less pronounced, and possibilities for improving the description of quantization effects are far from exhausted.
\section*{Acknowledgments}
The authors acknowledge computational resources provided by the HPC Core Facility and the HRZ of Justus Liebig University Giessen. We thank P. Risius and M. Giar of the HPC Core Facility for technical support and M. Czerner for advice concerning the KKR code.

\bibliography{references.bib}

\end{document}